\documentclass[aps,prd,showpacs]{revtex4}
\usepackage{graphicx}
\usepackage{amsmath,amsfonts,amssymb,amsthm,amstext,amscd,eucal}
\usepackage{bbold}
\usepackage{array}
\usepackage[latin1]{inputenc}

\newcommand{\eO}{\mathcal{O}}

\def\fun#1#2{\lower3.6pt\vbox{\baselineskip0pt\lineskip.9pt
  \ialign{$\mathsurround=0pt#1\hfil##\hfil$\crcr#2\crcr\sim\crcr}}}
\def\lap{\mathrel{\mathpalette\fun <}}
\def\gap{\mathrel{\mathpalette\fun >}}

\newcommand{\MUNCH}[1]{\relax}
\begin{document}

\title{Constraints on the redshift dependence of the dark energy potential} 
\author{Joan Simon}
\email{jsimon@bokchoy.hep.upenn.edu} 
\author{Licia Verde}
\email{lverde@physics.upenn.edu} 
\author{Raul Jimenez}
\email{raulj@physics.upenn.edu} 
\affiliation{ Dept. of Physics and
  Astronomy, University of Pennsylvania, 209 South 33rd Street,
  Philadelphia, PA-19104, USA}

\date{\today}

\begin{abstract}
  
  We develop a formalism to characterize the redshift evolution of the
  dark energy potential. Our formalism makes use of quantities similar
  to the Horizon-flow parameters in inflation and is general enough
  that can deal with multiscalar quintessence scenarios, exotic matter
  components, and higher order curvature corrections to General
  Relativity.  We show how the shape of the dark energy potential can
  be recovered non parametrically using this formalism and we present
  approximations analogous to the ones relevant to slow-roll
  inflation.  Since presently available data do not allow a
  non-parametric and exact reconstruction of the potential, we
  consider a general parametric description.  This reconstruction can
  also be used in other approaches followed in the literature (e.g.,
  the reconstruction of the redshift evolution of the dark energy
  equation of state $w(z)$).  Using observations of passively evolving
  galaxies and supernova data we derive constraints on the dark energy
  potential shape in the redshift range $0.1 < z < 1.8$. Our findings
  show that at the $1$$\sigma$ level the potential is consistent with
  being constant, although at the same level of confidence variations
  cannot be excluded with current data. We forecast constraints
  achievable with future data from the Atacama Cosmology Telescope.

\end{abstract}


\maketitle

\section{Introduction}
\label{sec:intro}

Recent observations \cite{Riess+04,SpergelWMAP03} indicate that
$\simeq$ 70\% of the present-day energy density of the universe may be
made of a dark energy component.  The two leading explanations of dark
energy are a cosmological constant or a slowly rolling scalar field
e.g.,\cite{PeeblesRatra,RatraPeebles,Caldwell98,Huey+99} but an
explanation in terms of modifications to the Friedman
equations(e.g.\cite{CDTT04,Freese+02} is also possible.  In both cases
this component has a negative pressure thus inducing an accelerated
expansion of the Universe.

A significant observational effort is directed to unveil the nature of
dark energy (e.g.,\footnote{WMAP: {\tt http://.map.gsfc.nasa.gov}},
\footnote{SNAP:{\tt http://snap.lbl.gov/}}, \footnote{DUO: {\tt
    http://duo.gsfc.nasa.gov/} }, \footnote{ACT:\cite{KosowskyACT}
  {\tt http://www.hep.upenn.edu/act/}}, \footnote{SPT:{\tt
    http://astro.uchicago.edu/spt/}}, \footnote{Planck:{\tt
    http://astro.estec.esa.nl/Planck/}}).  With few exceptions
\cite{Alam+04a,Alam+04b,Wang+04,Riess+04}, current constraints on the
nature of dark energy mostly measure the integrated value over time of
its equation of state parameter ($w=\rho/p$)
(e.g.,\cite{SpergelWMAP03,Riess+04,Kunz+04,Corasaniti04,TegmarkSDSS04})
or, alternatively its energy density as a function of time (e.g.,
\cite{WangTegmark04,WangFreese04}), which depends on the integral of
the equation of state parameter. These constraints are very tight
(e.g. \cite{SpergelWMAP03} finds $w=-0.98 \pm 0.12$) and are centered
around the expected value for the cosmological constant, but, as
pointed out by \cite{Maor,MaorBrunstein03}, the finding that the time
average value of $w$ is consistent with $-1$ does not exclude the
possibility that $w$ varied in time. Therefore, it is an open
challenge to determine whether dark energy is a cosmological constant
or a rolling of a scalar field.

A recent review of the current status of our knowledge of the
observational determination of $w$ and possible theoretical models to
explain it is given by \cite{PeeblesRatra}. 

From a theoretical point of view, it is not only important to clarify
whether this energy component is dynamical or constant, but, in case
it is not a cosmological constant, it is also of great interest to
constrain the potential of the rolling scalar field. Since different
theoretical models are typically characterised by different
potentials, a reconstruction of the dark energy potential from
observations can yield more direct constraints on physically motivated
dark energy models.

In this paper we present a non-parametric method to reconstruct the
redshift evolution of the potential and kinetic energy densities of
the dark energy field. Our formalism introduces quantities similar to
the Horizon-flow parameters \cite{schwarz,llms} in inflation.  It has
the nice feature that it is easily implemented in the presence of
higher order curvature corrections to General Relativity and different
types of energy contributions in Einstein's equations, as we do in
Section \ref{sec:dyn}.  Our exact reconstruction formulas determine
the value of the potential at a given redshift once the matter
density, Hubble parameter $(H)$ and its first derivative $(\dot{H})$
are experimentally measured at that redshift value.  We discuss the
observational challenges to reconstruct the potential in this fully
non-parametric way due to the difficulty in measuring $\dot{H}$.  As
current data is not good enough to determine $\dot{H}$, we present a
general parameterization of the potential, based on an expansion in
Chebyshev polynomials. In this approach, the scalar potential function
at a given redshift is expanded in Chebyshev polynomials, which
constitutes a complete orthonormal basis on a finite interval, and
have the nice property to be the {\em minimax} approximating
polynomial. Our reconstruction equation becomes a differential
equation for the Hubble parameter, which we solve analytically, and
the coefficients in the Chebyshev expansion become the parameters to
be constrained from observations of the Hubble parameter.  Our general
parameterization can apply to other approaches that were already
considered in the literature, such as expansions of the equation of
state and we show the correspondence to some parameterizations that
have been proposed in the literature.  Using current data (in
particular with recent supernovae data and relative ages of a sample
of passively evolving galaxies) we reconstruct the potential of dark
energy using our parameterization up to $z \sim 1.8$.  The
reconstructed potentials obtained from galaxy ages and SN are
consistent. Since these two data sets rely on independent physics and
are affected by completely different systematics, this finding
suggests that possible systematics are not a crucial issue.

The reconstructed potential is consistent with being constant up to
the maximum redshift of the observations, although current constraints
do not exclude a variation as a function of redshift. We show that
data obtained with the Atacama Cosmology Telescope will be able to
greatly improve current constraints.

\section{Method}

\subsection{Dynamics of the scalar field of dark energy}
 \label{sec:dyn}
 
 The classical effective action that we shall use to describe the
 dynamics of the universe is
\begin{equation}
  S = \int\,dt\,d^3 x\,\sqrt{-g}\left\{-\frac{m_p^2}{16\pi}\,\left(R
  + f(R,\, R^{\mu\nu}R_{\mu\nu},\dots\right) +
  \frac{g^{\mu\nu}}{2}\,\partial_\mu q\,\partial_\nu q -
  V(q)\right\}
  + S_{\text{sources}}\,,
 \label{eq:actiona}
\end{equation}
where $m_p$ stands for the four dimensional Planck mass and $g_{\mu\nu}$
for the components of the four dimensional metric, 
\begin{equation}
  ds^2 = dt^2 - a^2(t)\,d{\bf x}^2~,
\label{eq:metricd4}
\end{equation}
which we shall consider to be an homogeneous, isotropic and spatially
flat FRW cosmology, as supported by recent data \cite{SpergelWMAP03}.

$S_{\text{sources}}$ stands for the classical action describing the
physical energy content, such as matter and radiation, but it could
also include more exotic sources (e.g. defects, cosmic strings etc.).
Note also that we have implicitly assumed the existence of a single
canonically normalised quintessence scalar field $q(t,\,\vec{x})$
subject to the potential $V(q)$. Thus, we have assumed that this
potential is independent of the derivatives of the scalar field.  For
generality, in Eq. \eqref{eq:actiona} we include the effect of higher
derivative terms in the gravitational sector of the theory
\cite{CDTT04}. These are described by the function
$f(R,\,R^{\mu\nu}R_{\mu\nu},\dots)$ of the different invariants that
we can construct out of the metric and its derivatives. In four
dimensions, the most general lowest order corrections to Einstein's
classical action would be described by 
$f=\beta\,R^2 + \delta\,R^{\mu\nu}R_{\mu\nu}\,$  
\footnote{These are corrections
  that are known to be generated by quantum corrections to the
  classical action, and in particular, the ones considered here
  involve a linear combination of the set of independent operators of
  lowest dimension.}
(see, for example, \cite{birrelldavies}).  Other corrections that have
been considered, include arbitrary functions of the scalar curvature
$f(R)$ \cite{review}, which include as particular examples linear
combinations of negative powers of these invariants \cite{seanmark}.

We focus on cosmologies given by \eqref{eq:metricd4}, and 
shall restrict ourselves to classical configurations $q=q(t)$,
configurations that do not break the homogeneity and isotropy of spacetime.
The energy momentum tensor of this scalar field configuration is that of a
perfect fluid, with density $\rho_q$ and pressure $p_q$ given by
\begin{equation}
  \rho_q = K(q)+ V(q)\,, \quad
  p_q = K(q)- V(q)\, \quad \mbox{and} \,\,\,\, K\equiv\frac{1}{2}\dot{q}^2 .
 \label{eq:qtensor}
\end{equation}
where $K$ denotes the kinetic energy of the field.
Under these assumptions, one is led to consider Einstein's equations,
plus the Klein-Gordon equation of motion for the scalar field. The
first ones reduce to Friedmann's equations
\begin{equation}
  \begin{aligned}[m]
    H^2 & = \frac{\kappa}{3}\,\left(\rho_T + \rho_q\right)\,, \\
    \frac{\ddot{a}}{a} &= - \frac{\kappa}{6}\left(\rho_T + 3p_T 
    + \rho_q + 3\,p_q\right)~,
  \end{aligned}
 \label{eq:friedmann}
\end{equation}
where $\kappa = 8\pi/m_p^2$ (or $\kappa = 8\pi G$). In Eq. \eqref{eq:friedmann} we introduced
the compact notation $\rho_T$ and $p_T$ for the total energy density
and pressure. For example $\rho_T$  denotes the full energy density
contribution of $S_{\text{sources}}$ and of the higher derivative curvature
terms $f(R,\,R^{\mu\nu}R_{\mu\nu},\dots)$. Thus if the sources are a collection 
of $n$ perfect fluids with constant equation of state $\omega_i$ $i=1,\dots,n$, 
$\rho_T$ and $p_T$  are
\begin{equation}
  \rho_T = \sum_{i=1}^n \rho_i + \rho_f\,,\quad
  p_T = \sum_{i=1}^n \omega_i\,\rho_i + p_f~,
\end{equation}
where $\rho_f$ and $p_f$ describe the contribution from the
higher derivative curvature terms. For the particular function $f(R^2,\,R^{\mu\nu}R_{\mu\nu})$
introduced above, such terms would be written as
\begin{eqnarray*}
  \kappa\,\rho_f &=& -12H\,\frac{a^{(3)}}{a}\,\left(3\beta+\delta\right) -
  6 \left(\frac{\ddot{a}}{a}\right)^2\,\left(3\beta-\delta\right) 
  + 6H^4\,\left(15\beta+7\delta\right) 
  - 36H^2\,\frac{\ddot{a}}{a}\,\left(\beta+\delta\right) \\
  \kappa\,p_f  &=& 4\frac{a^{(4)}}{a}\,\left(\delta+3\beta\right) 
  + 12H\,\frac{a^{(3)}}{a}\,\left(\delta+2\beta\right) 
  + 8H^2\,\frac{\ddot{a}}{a}\,\left(\delta+9\beta\right)
  +2\left(\frac{\ddot{a}}{a}\right)^2\,\left(\delta+3\beta\right)
  + 2H^4\,\left(15\beta + 7\delta\right)~.
\end{eqnarray*}

On the other hand, the scalar field $q(t)$ equation of motion reduces
to
\begin{equation}
  \ddot{q} + 3H\,\dot{q} + V^\prime = 0~,
 \label{eq:qmotion}
\end{equation}
where $V^\prime= dV/dq$.

\subsection{Reconstruction procedure}
 \label{sec:rcons}
 
 In this section we provide exact analytical expressions in which
 both, the kinetic and potential energies of the quintessence field
 $q(t)$, depend on quantities more directly observable such as the
 energy densities, the Hubble constant $H$ and its derivatives.
 Although the higher curvature corrections are not directly
 observable, they will also have to appear in the expressions: they
 can be taken into account for a given model that is for a given
 parameterization of the functional $f$.  Provided one has an
 independent way of determining the densities, $H$ and $\dot{H}$, the
 value of the potential $V(z)$ at a given redshift $z$ where these
 measurements are available, can then be fixed, up to experimental
 uncertainties. If also higher order derivatives of $H$ are known,
 higher derivatives of the potential can be determined
 $\frac{d^{(s)}V}{dq^{(s)}}(z)$, which can be used to probe the
 flatness of the potential.
 
 We use the analogous of the inflationary {\it horizon-flow
   parameters} \cite{schwarz,llms} $\{\varepsilon_n\}$, which are
 defined recursively by
\[
  \varepsilon_{n+1} = \frac{d\log\mid\varepsilon_n\mid}{dN}\,, \quad
  n\geq 0
\]
where $N=\log (a(t)/a(t_i))$ is the number of e-foldings since some
initial time $t_i$ and $\varepsilon_0 = H(N_i)/H(N)$.  There are many
similarities between the period of inflation and the present-day
accelerated expansion, but, despite the fact that inflation happened
13.7 billion years ago, and the accelerated expansion is happening
today, as we will see, it is not observationally easier to reconstruct
the dark energy potential than it is to reconstruct the inflationary
potential.  In the equation describing inflationary dynamics the
contribution due to matter can be ignored, but it can't be ignored
when describing today's expansion. Moreover, the detailed shape of the
primordial power spectum from CMB scales to large scale structure
scales, and the nature of the primordial perturbations offer a window
to test the last 4 inflation efoldings; conversely, in the case of
dark energy, dark energy started dominating at $z < 1$, and between
then and now the Universe expanded only by a factor $<2$. In addition
we can measure with exquisite precision perturbations from inflation
but have not detected perturbations from dark energy, which
is a very challenging task \cite{BeanDore04}. On the other hand we do
not have strong constraints on the energy scale of inflation, that is
on the ``normalization'' of the inflationary potential
\cite{PeirisWMAP03,Kinney+04} but as we will see, since the matter
content of the Universe can be independently determined, for a flat
Universe, we have some constraints on the quintessence potential
normalization. 

Keeping in mind the different kind of challenges that a quintessence
potential reconstruction faces, we proceed with our program. For our
purposes, it will be useful to have explicit expressions for the first
two parameters
\begin{equation}
  \varepsilon_1 = -\frac{\dot{H}}{H^2} = 1-
  \frac{\ddot{a}}{a}\,H^{-2} = \frac{dH}{dz}\frac{(1+z)}{H}
\end{equation}

\begin{equation}
\varepsilon_2 =
  \frac{\dot{\varepsilon}_1}{H\,\varepsilon_1}~,
 \label{eq:horizonpar}
\end{equation}

which, we will show, are needed to determine $V$ and
$V'$ \footnote{Throughout this paper we denote with ' the derivative
  with respect to q and with $\dot{ }$ derivative with respect to time.}.  
We use the second Friedmann equation \eqref{eq:friedmann} to
express the first horizon-flow parameter $\varepsilon_1$ in terms of
the energy and pressure densities:
\begin{equation}
  \varepsilon_1 =
  \frac{3}{2}\,\frac{\rho_T +\rho_q + p_T + p_q}{\rho_T +\rho_q}~.
 \label{eq:horizona}
\end{equation}

If we write $\{\rho_q,\,p_q\}$ in terms of its kinetic and potential
energy components, as in \eqref{eq:qtensor}, we can use
\eqref{eq:horizona} to express, e.g., the kinetic energy in terms of
the potential energy as 
\begin{equation}
  \frac{1}{2}\dot{q}^2 = \frac{1}{3-\varepsilon_1}\,\left\{
  \varepsilon_1\left(\rho_T + V\right) -\frac{3}{2}\left(\rho_T +
  p_T\right)\right\}~.
 \label{eq:qdot}
\end{equation}

Finally we can use the first
Friedmann equation \eqref{eq:friedmann} to solve for the value of the
kinetic energy and the potential at a given redshift $z$
\begin{equation}
  K(z) = \frac{1}{2}\dot{q}^2 = \varepsilon_1\,\frac{H^2}{\kappa} - \frac{1}{2}(\rho_T+p_T)~,
 \label{eq:kinetic}
\end{equation}
\begin{equation}
  V(z) = (3-\varepsilon_1)\,\frac{H^2}{\kappa}
  + \frac{1}{2}\left(p_T-\rho_T\right)~.
 \label{eq:potential}
\end{equation}
This is the generalisation of eq. (16) in \cite{llms} which was derived in the
context of inflation.  

Equation \eqref{eq:potential} is a  general and exact reconstruction
formula for the potential of a quintessence field given the
assumptions followed in this paper. 

Here, we shall focus in the constraints on
the potential at redshifts smaller than $1000$. Therefore, we shall neglect the radiation
energy density contribution. Furthermore, if we also
neglect the contribution from the higher-order curvature terms, the
expression for the potential simplifies
\begin{equation}
  V(z) = (3-\varepsilon_1)\,\frac{H^2}{\kappa} -\frac{1}{2}\,\rho_m~.
 \label{eq:rpotential}
\end{equation}
Analogously, for the  kinetic energy we obtain 
\begin{equation}
  K(z) = \varepsilon_1\,\frac{H^2}{\kappa} -\frac{1}{2}\rho_m~.
 \label{eq.rkinetic}
\end{equation}

Ideally, our goal would be  to constrain the functional form of the
potential, $V[q]$, and this is not what \eqref{eq:potential} provides. 
If the function $q(z)$ was known this would be straightforward, but
$q(z)$ is {\it not} an observable quantity.
We will show later that $V[q]$ can be obtained if equation
(\ref{eq.rkinetic}) can be integrated. 

We can next determine the first derivative of the quintessence
potential. We rewrite \eqref{eq:qmotion} as
\begin{equation}
  V^\prime = -(\dot{q})^{-1}\left\{3H\dot{q}^2 +
  \dot{q}\,\ddot{q}\right\}~.
 \label{eq:vprimef}
\end{equation}
where all terms are already known, except for $\dot{q}\,\ddot{q}$ which can
be  obtained from the time derivative of the
kinetic energy \eqref{eq:kinetic}. The end result can be expressed as
\begin{multline}
  V^\prime = -3\frac{m_p}{\sqrt{4\pi}}\,H^2\,(\varepsilon_1)^{1/2}
  \left\{1-\frac{\kappa}{2\,H^2\,\varepsilon_1}\,[\rho_T +
  p_T]\right\}^{-1/2}
  \left\{1+\frac{\varepsilon_2}{6}-\frac{\varepsilon_1}{3}
  - \frac{\kappa}{6\,H^3\,\varepsilon_1}\left[3H\,(\rho_T+p_T) +
  \frac{1}{2}\left(\dot{\rho}_T + \dot{p}_T\right)\right]\right\}~.
 \label{eq:vprime}
\end{multline}

Thus, if the values of  $\rho_T(z),p_T(z)$ and
$\{H,\,\varepsilon_1,\,\varepsilon_2\}$ or equivalently,
$\{H,\,\dot{H},\,\ddot{H}\}$, can be experimentally determined for
some redshift $z$,  \eqref{eq:vprime} yields the first derivative of the potential
$V^\prime(z)$. As in the previous
discussion, the determination of $V^\prime[q(z)]$ would require the
knowledge of $q(z)$.

The above formula is the exact result given some energy density
content $\rho_T$, with associated pressure $p_T$. 
If we restrict ourselves to a single matter component and neglect the
higher order curvature terms, the first potential derivative reduces
to
\begin{equation}
  V^\prime(z) = -3\frac{m_p}{\sqrt{4\pi}}\,H^2\,(\varepsilon_1)^{1/2}
  \left\{1-\frac{\kappa}{2\,H^2\,\varepsilon_1}\,\rho_m\right\}^{-1/2} 
  \left\{1+\frac{\varepsilon_2}{6}-\frac{\varepsilon_1}{3}
  - \frac{\kappa}{4\,H^3\,\varepsilon_1}\,\rho_m\right\}~.
 \label{eq:rvprime}
\end{equation}
and \eqref{eq:vprimef} reproduces equation (17) in \cite{llms}
when the matter density vanishes $(\rho_m=0)$.

In this case, the first derivative of the potential
$V^\prime(z)$ is known if one can measure  $\rho_m(z=0)$, $\{H,\,\dot{H},\,\ddot{H}\}$.

Analogously exact
expressions for higher order derivatives of the potential
$d^{(r)}V[q]/dq^r$ can be obtained by taking the time derivative of
\eqref{eq:vprime}  and using the exact expression for the kinetic
energy \eqref{eq:kinetic}. 

\subsection{Redshift parameterisation of the potential}
\label{sec:redpar}

In section \ref{sec:rcons} we have shown that an exact reconstrution
of $V(z)$ is possible only if $H(z)$ and $\dot{H}(z)$ are known. While
the determination $H(z)$ is an observationally challenging task (e.g,
\cite{JVTS03,DalyD03} and \ref{sec:exp}), the determination of
$\dot{H}(z)$ is even more formidable.  In this section we shall not
attempt a non-parametric and exact reconstruction of $V(z)$, we shall
instead consider a parametric description of the potential
$(V(\alpha_i,\,z))$ in terms of the redshift $z$ and parameters
$\alpha_i$. In section \ref{sec:exp} we will then use currently
available observations to constrain the potential parameters and
discuss future prospects. Hereafter we will set $\rho_f \equiv 0$ and
defer the more general case of $\rho_f \ne 0$ to future work.

Equation
\eqref{eq:rpotential}, can be  rewritten in terms of the independent
variable $z$ as
\begin{equation}
  3H^2(z) - \frac{1}{2}\,(1+z)\,\frac{d\,H^2(z)}{dz} =
  \kappa\,\left(V(\alpha_i,\,z) + \frac{1}{2}\rho_m(z)\right) \equiv
  g(\alpha_i,\,z)~.
 \label{eq:Hz}
\end{equation}
This is a first order non-linear differential equation which can
be integrated analytically:
\begin{eqnarray}
  H^2(\alpha_i,\,z) &=& H_0^2 (1+z)^6 - 2(1+z)^6\,\int_0^z\,
  g(\alpha_i,\,x)\,(1+x)^{-7}\,dx~\\ \nonumber
  &=&  \left(H_0^2-\frac{\kappa}{3} \rho_{m,0}\right)\,(1+z)^6 +
  \frac{\kappa}{3} \rho_m(z)
  -2(1+z)^6\int_0^z\,V(\alpha_i,x)\,(1+x)^{-7}\,dx~.
 \label{eq:Hzsol}
\end{eqnarray}
Hereafter the $0$ subscript denotes the quantity evaluated at $z=0$.

In this approach if we now consider the kinetic energy of the
quintessence field we obtain a first-order non-linear differential
equation for $q(z)$
\begin{equation}
  \frac{1}{2}\dot(q)^2=(1+z)^6V_0-6(1+z)^6\int_0^zV(\alpha_i,z)(1+z)^{-7}dz
  + K_0 (1+z)^6\,.
\end{equation}
or equivalently

\begin{equation}
  \frac{1}{2}\left(\frac{dq}{dz}\right)^2\,(1+z)^2\,H^2(\alpha_i,\,z)
  = 3\kappa^{-1}\,H^2(\alpha_i,\,z) - \rho_m(z) - V(\alpha_i,\,z)~,
 \label{eq:qz}
\end{equation}
which can be integrated to obtain $q(z)$ and thus $V[\alpha_i,\,q]$
from $V(\alpha_i,\,z)$:

\begin{equation}
  q(z)-q(0) = \pm \int^z_0\,
  \frac{dz}{(1+z)\,H(\alpha_i,\,z)}\,\left\{6\kappa^{-1}\,
  H^2(\alpha_i,\,z) - 2\rho_m(z) - 2V(\alpha_i,\,z)\right\}^{1/2}~,
 \label{eq:qzsol}
\end{equation}
where the ambiguity in sign comes from the quadratic expression for the kinetic
energy. Typically, if we think of an scalar field rolling slowly along its potential, the plus sign will be the relevant one.

For example let's consider a simple  two-parameters parameterization of the potential:
\begin{equation}
  V = \lambda\,(1+z)^\alpha~,
\label{eq:ratra-peebles_Vz}
\end{equation}
which yields
\begin{equation}
  H^2(z) = H_0(1+z)^6 - 2I_\alpha\,(1+z)^6~,
\end{equation}
where
\begin{eqnarray}
  I_\alpha &=& -\frac{\kappa}{6}\rho_m(0)\,\left[(1+z)^{-3}-1\right] + 
  \frac{\lambda\,\kappa}{\alpha-6}\,\left[(1+z)^{\alpha-6}-1\right]~,
  \quad \alpha\neq 6 \\
  I_6 &=& -\frac{\kappa}{6}\rho_m(0)\,\left[(1+z)^{-3}-1\right] + 
  \lambda\,\kappa\log (1+z)~, \quad \alpha=6
\end{eqnarray}

If we can neglect the kinetic energy (that is if $\alpha <<1$)  then this potential correspond to
a constant equation of state $w$: $\alpha=3(w+1)$ as in the Ratra-Peebles
\cite{PeeblesRatra} case and $\Omega_{q,0}=6\lambda/[\rho_c(6-\alpha)]$.

\subsection{Chebyshev reconstruction}
\label{sec.V_cheby}

An interesting parameterization of the potential involves the
Chebyshev polynomials, which form a complete set of orthonormal functions
on the interval $[-1,1]$. They also have the interesting property to be 
the minimax approximating polynomial, that is, the approximating polynomial 
which has the smallest maximum deviation from the true function at any given 
order. We can thus approximate a generic $V(z)$ as
\begin{equation}
V(z)\simeq\sum_{n=0}^{N}\lambda_n T_n(x)
\label{eq:V-cheb-expan}
\end{equation}
where $T_n$ denotes the Chebyshev polynomial of order $n$ and  we have normalized the redshift interval so that
$x=2z/z_{max}-1$; $z_{max}$ is the maximum redshift at which
observations are available and thus $x\in [-1,1]$. Since $|T_n(x)| \le
1$ for all $n$, for most applications, an estimate of the error
introduced by this approximation is given by $\lambda_{N+1}$.  With
this parameterization, the relevant integral in (\ref{eq:Hzsol})
becomes:
\begin{equation}
  \int_0^zV(y)(1+y)^{-7}dy=\frac{z_{max}}{2}\sum_{n=0}^N\lambda_n\int_{-1}^{2z/z_{max}-1}
  T_n(x)(a+bx)^{-7}dx \equiv \sum_{n=1}^{N}\lambda_n\,F_n(z)
 \label{eq:chebxzpn}
\end{equation}
where $a=1+z_{max}/2$ and $b=z_{max}/2$.
These integrals can be solved analytically for any order $n$ as shown
in the appendix: $F_n$ are known analytic functions which are reported in appendix A.
 
We obtain:
\begin{equation}
H^2(z,\lambda_i)=(1+z)^6 H_0^2\left[1 -3 z_{max} \sum_{n=0}^N
  \frac{\lambda_n}{\rho_{c}} F_n(z)-\Omega_{m,0}\left(1-\frac{1}{(1+z)^3} \right)\right]
\label{eq:Hsq-cheby}
\end{equation}

where $\rho_c$ denotes the present-day critical density.

Equation (\ref{eq:Hsq-cheby}) seems to describe the potential with
$N+1$ parameters ($\lambda_0 \cdot \cdot \cdot \lambda_N$). However,
since we assume a flat Universe, $\Omega_{m,0}$ constrains the
coefficients of the Chebyshev polynomials in expansion
(\ref{eq:V-cheb-expan}) and the kinetic energy of the field.  For
example, if the potential is constant (i.e in a cosmological constant
case) then it is completely described by only one parameter,
$\lambda_0$, and since $\dot{q}=0$,
$V_0+K_0=V_0=\rho_{q,0}=\Omega_{q,0}\rho_c$, we have that
$\lambda_0=\Omega_{q,0}\rho_c$.  However if the potential is not
constant we obtain

\begin{eqnarray}
\frac{1}{2}\dot{q}^2&=&V_0(1+z)^6-V(z)-6(1+z)^6\int_0^zV(z)(1+z)^{-7}dz+K_0(1+z)^6\\
\nonumber
&=&V_0(1+z)^6-V(z)-3(1+z)^6z_{max}\sum_n\lambda_n F_n+K_0(1+z)^6\,.
\label{eq.kinetic_cheby}
\end{eqnarray}
and since $V_0=\sum_{i=0}^N\lambda_i(-1)^n$ we have the constraint:
\begin{equation}
\sum_{i=0}^N\lambda_i(-1)^n+K_0=\Omega_{q,0}\rho_c\,.
\end{equation}

In section \ref{sec:exp} we show the constraints that can be obtained for the
first few Chebyshev coefficients from currently available data.

\section{Equation of state}

It is widespread to parameterize dark energy not by the scalar field
potential but by its equation of state. In this section we connect the two descriptions. 

\subsection{Time dependent density scaling}
 \label{sec:scaling}
 
 Standard contributions to the energy momentum tensor in Einstein's
 equations are characterized by a parameter that governs how their
 energy densities decrease with the expansion of the universe. For the
 energy density of the scalar field we can write in all generality
\begin{equation}
  \rho_q(t) = \rho_q(0)\left(\frac{a_0}{a(t)}\right)^{\gamma(t)}~.
 \label{eq:qscaling}
\end{equation}

The Klein-Gordon equation \eqref{eq:qmotion} can be expressed as the
conservation equation for the energy momentum tensor describing the
scalar field,
\begin{equation}
  \dot{\rho_q} + 3H\,\left(\rho_q + p_q\right) =0~.
 \label{eq:qlaw}
\end{equation}
Thus, the pressure of the scalar field $p_q(t)$ can be expressed as a
function of the time-dependent exponent $\gamma(t)$
\begin{equation}
  p_q(t) = \left[\frac{\gamma (t)-3}{3} - \frac{1}{3}
  \log\left(\frac{a_0}{a(t)}\right)\,H^{-1}\,\frac{d\gamma}{dt}\right]\,
  \rho_q(t)~.
\end{equation}
Using ansatz \eqref{eq:qscaling}, we
obtain the kinetic and potential energies for the scalar field
\begin{eqnarray}
  \frac{1}{2}\dot{q}^2 &=& \frac{1}{2}\,\rho_q(t)\,\Delta w_q~, \\
  V &=& \rho_q\,\left(1-\frac{1}{2}\,\Delta w_q(t)\right)~,
 \label{eq:scalepotential}
\end{eqnarray}
where we introduced the function (compare to \cite{Bludman04})
\begin{equation}
  \Delta w_q(t) = \frac{1}{3}\left[\gamma(t) - 
  H^{-1}\,\log\left(\frac{a_0}{a(t)}\right)\,
  \frac{d\gamma}{dt}\right]~
 \label{eq:F(t)}
\end{equation}
which depens on the  ratio  between kinetic and
potential energies of the scalar field via
\begin{equation}
  \Delta w_q(t) = \frac{2}{1+V[q]/K[q]}~.
 \label{eq:obsF(t)}
\end{equation}

The function 
$\Delta w_q(t)$ controls the deviations from the equation of state
of the scalar field $w_q(t)$ from being exactly $-1$ i.e. a
cosmological constant. Indeed, from 
$p_q(t)=w_q(t)\,\rho_q(t)$, we have 
\begin{equation}
  w_q(t) = -1 + \Delta w_q(t)~.
 \label{eq:eqstate}
\end{equation}
Thus constraining $\Delta w_q(t)$ is equivalent to constraining
the time (or redshift) evolution of the dark energy equation of state
for which there are different independent observational constraints
(e.g., \cite{PeeblesRatra,hu04}, and references therein)

We can relate $\gamma(t)$ to $\Delta w_q(t)$  by integrating \eqref{eq:F(t)}.
Without loosing generality, we do this in terms of the redshift
\begin{equation}
  \gamma (z)= \frac{3}{\log\,(1+z)}\,\int_{0}^{z}\,\Delta w_q(y)\,\frac{dy}{1+y}~,
\end{equation}
where we did not include a possible integration constant, since it is physically irrelevant, 
i.e. it just redefines the value of the energy density of the scalar field today $(\rho_q(0))$.
This relation provides an expression for the potential energy density of the scalar field 
once the equation of state is known, using \eqref{eq:scalepotential}:
\begin{equation}
  V(z) = \rho_q(0)\,e^{3\int_{0}^{z}\,\Delta w_q(y)\,\frac{dy}{1+y}}\,
  \left(1-\frac{1}{2}\Delta w_q(z)\right)~.
 \label{eq:potvsomega}
\end{equation}
The case of a constant $\Delta w_q$, or equivalently, a constant equation of state $w_q$
would correspond to the redshift parameterization considered in \eqref{eq:ratra-peebles_Vz}
with 
\begin{equation*}
  \lambda=\rho_q(0)\,\left(1-\frac{1}{2}\Delta w_q\right)~, \quad
  \text{and} \quad \alpha = 3(1+w_q)~.
\end{equation*}
Thus, for a physical situation resembling a cosmological constant
$(w_q\sim -1)$, the parameter $\alpha\ll 1$, as we claimed in
the previous section. Some constant scalar field energy density
scalings were discussed by \cite{RatraPeebles}. The
above formalism is a natural generalisation of their models.

\subsection{Beyond a rolling scalar field }
 \label{sec.w_cheby}

Up to now we have assumed that dark energy is given by a scalar
field. The techniques developed here however can also be used in a
more general context. 

If the dark energy is not due to a scalar field we can still describe
it as a fluid with a given equation of state.  As long as the equation
of state $w$ is $>-1$ the two descriptions are equivalent.  For
example for a given set of parameters $\lambda_i$ and $H_0$ and
$\Omega_{m,0}$, $V(\lambda_i,z)$ is known, the kinetic energy can be
computed using eq. \ref{eq.kinetic_cheby}, thus $\Delta w$ is known
(eq. \ref{eq:obsF(t)}) and so $w(z)$ (eq. \ref{eq:eqstate}).

However if we want to allow $w<-1$ then the scalar field
description as presented here fails. However, analogously to section
\ref{sec.V_cheby}, we can expand the redshift
dependence of $w$ in Chebyshev polynomials without imposing any
restrictions on the values that $w$ can take. We thus obtain
\begin{equation}
w(z) \simeq \sum_{i=0}^{N}\omega_iT(x(z))
\label{eq:w_cheby}
\end{equation}
and 
\begin{equation}
  H^2(\omega_i,z)\simeq H_0^2(1+z)^3\left[\Omega_{m,0}+\Omega_{q,0}\exp\left(\frac{3}{2}z_{max}\sum_{n=0}^N
  \omega_n G_n(z)\right) \right]
 \label{eq:Hzwcheby}
\end{equation}

where $G_n$ is the analogous of $F_n$ of section \ref{sec.V_cheby}: $F_n$ is a
linear combination of integrals $I_i$, $i=\{0,n\}$ while $G_n$ is the
same linear combination of the integrals $J_i$ , $i=\{0,n\}$ (see appendix).

Note that in this parameterization the present-day value of $w$ is
given by
\begin{equation}
w_0=\sum_{i=0}^N(-1)^i\omega_i
\label{eq:wtoday}
\end{equation}

 Given the  parameterization (\ref{eq:w_cheby}) of $w(z)$ subject
 to the constraint $w>-1$ one can always obtain:
\begin{equation}
V(z)=\rho_{q,0}(1+z)^{3(1+\omega_0)}\exp\left[\frac{3}{2}z_{max}\sum_{i=1}^N
  \omega_i G_i(z)\right]\frac{1}{2}\left[1-\sum_{i=0}^N \omega_i T_i(x(z))\right]
\end{equation}

In section \ref{sec:exp} we show how currently available data can be
used to constrain the first few Chebyshev coefficients of this expansion. 
 
In the remaining of this section we will compare some models
presented in the literature with the parameterization presented here.
Clearly, the case of a constant equation of state corresponds to
 $\omega_i=0$ for $i>0$. The linear parameterization in $z$
\cite{HutererTurner01, WellerAlbrech02} corresponds to    $\omega_i=0$
 for $i>1$, and in particular $w_0=\omega0-\omega_1$ and $w'=2
 \omega_1/z_{max}$. Finally 
the linear parameterization in $a$
\cite{ChevallierPolarskiStarobinsky01,Linder03},
 $w=w_0+w_a z/(1+z)$  for $w_a \ll w_0$ can be closely approximated
 by  $\omega_i=0$ for $i>2$, with the constraint (\ref{eq:wtoday}).
\cite{Bassett+04} pointed out that a simple, 2-parameter fit may
introduce biases: the expansion (\ref{eq:w_cheby}) allows one to
include more parameters  by increasing $N$ as the observational data 
improve. 

\section{Observational determination of $H(z)$}
\label{sec:exp}

Section \ref{sec:rcons} has illustrated that it is necessary to
determine observationally $H(z), \dot H(z), \ddot H(z)$ in order to
reconstruct $V[q]$ and its first derivative $V^\prime[q]$. 
Here we present a determination of $H(z)$ based on
the method developed by \cite{JL02} and we emphasize the difficulties
of computing $\dot H(z), \ddot H(z)$. We also present the constraints
that can be achieved on the evolution of the quintessence potential
and the dark energy equation of state from present and future data.
   
\subsection{Differential ages of passively evolving galaxies}

The Hubble parameter depends on the differential age of the universe
as a function of redshift in the form

\begin{equation}
H(z)=- \frac{1}{1+z} \frac{dz}{dt}.
\label{eq:hz}
\end{equation}

Therefore a determination of $dz/dt$ directly measures $H(z)$.
In \cite{JVTS03} we demonstrated the feasibility of the method by
applying it to a $z \sim 0$ sample. In particular, we used the Sloan
Digital Sky Survey to determine $H(0)$ and showed that its value is in
good agreement with other independent methods (see \cite{JVTS03} for
more details). With the availability of new galaxy surveys it becomes
possible to determine $H(z)$ at $z > 0$. Here we use the new publicly
released GDDS survey \cite{Abraham+04} and archival data
\cite{TSCMB99,TSMCB01,TSCMB02,D+96,Spinrad+97,NDJH03} to determine
$H(z)$ in the redshift range $0.1 < z < 1.8$.  We proceed as follows:
first we select galaxy samples of passively evolving galaxies with
high-quality spectroscopy.  Second, we use synthetic stellar
population models to constrain the age of the oldest stars in the
galaxy (after marginalising over the metallicity and star formation
history), in similar fashion as is done in \cite{JVTS03}. We compute
{\em differential} ages and use them as our estimator for $dz/dt$,
which in turn gives $H(z)$.

The first sample is composed of field early-type galaxies from
\cite{TSCMB99,TSMCB01,TSCMB02}. In \cite{JVTS03} we derived ages for
this sample using the SPEED models \cite{JMDPP04}. The second sample
is from the publicly released Gemini Deep Survey
(GDDS)\cite{Abraham+04}. GDDS has high-quality spectroscopy of red
galaxies, some of which show stellar absorption features, indicating
an old stellar population. The GDDS collaboration has determined ages
(and the star formation history) for these galaxies
\cite{McCarthy+04}: they conclude that for a sub-sample of 20 red
galaxies the most likely star formation history is that of a single
burst of star formation of duration less than $0.1$ Gyr (in most cases
the duration of the burst is consistent with $0$ Gyr, i.e. the
galaxies have been evolving passively since their initial burst of
star formation). To determine the galaxies ages they use a set of
stellar population models different than SPEED.  We have re-analize
the GDDS old sample using SPEED models and obtained ages within $0.1$
Gyr of the GDDS collaboration estimate. This indicates that
systematics are not a serious source of error for these high-redshift
galaxies. We complete our data set by adding the two radio galaxies
53W091 and 53W069 \cite{D+96,Spinrad+97,NDJH03}. In total we have $32$
galaxies.

\begin{figure}
\includegraphics[scale=0.8]{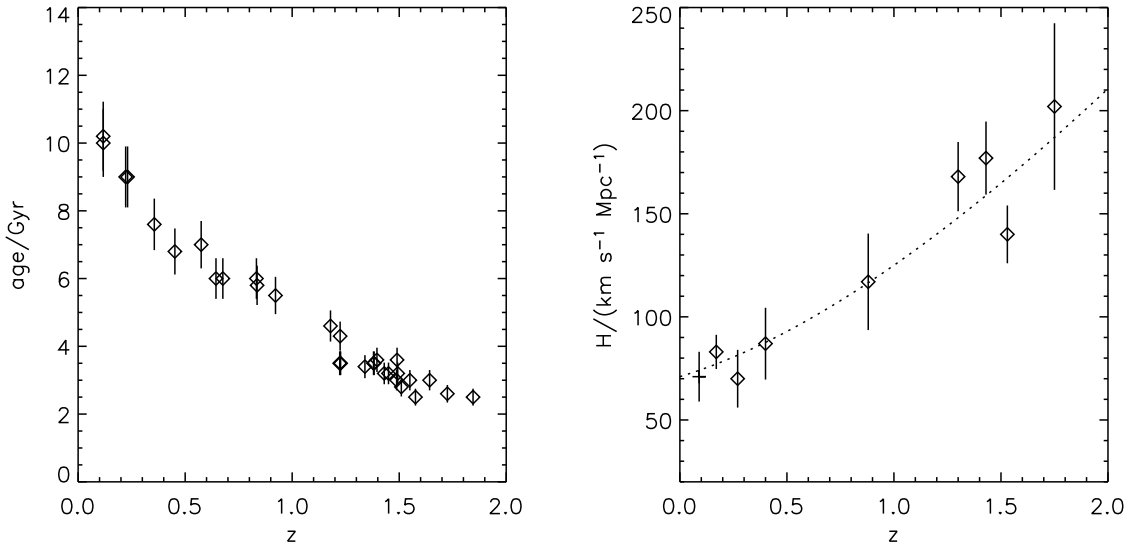}
\caption{Left panel: the absolute age for the 32 passively evolving galaxies in our catalogue (see text for more detals) determined from fitting stellar 
  population models is plotted as a function of redshift. Note that
  there is a clear age-redshift relation: the lower the redshift the
  older the galaxies. Right panel: the value of the Hubble parameter
  as a function of redshift as derived from the {\em differential}
  ages of galaxies in the left panel. The determination at $z\sim 0.1$
  indicated by the '+' symbol is the hubble constant determination of
  $H$ from \cite{JVTS03}. The dotted line is the value of
  $H(z)$ for the LCDM model.}
\label{fig:agevsz}
\end{figure}

Fig.~\ref{fig:agevsz} (left panel) shows the estimated {\em absolute}
ages for galaxies in the above samples and their $1\sigma$ error bars.
There is a distinguishable ``red envelope'': galaxies are older at
lower redshifts.

The next step is to compute differential ages at different redshifts
from this sample. To do so we proceed as follows: first we group
together all galaxies that are within $\Delta z=0.03$ of each other.
This gives an estimate of the age of the universe at a given redshift
with as many galaxies as possible. The interval in redshift is small
to avoid incorporating galaxies that have already evolved in age, 
but large enough for our sparse sample to have more than one galaxy in most of the bins. We then compute age
differences only for those bins in redshift that are separated more
than $\Delta z= 0.1$ but no more than $\Delta z = 0.15$. The first
limit is imposed so that the age evolution between the two bins is
larger than the error in the age determination.  
This provides with a robust determination
of $dz/dt$. We note here that differential ages are less sensitive to
systematics errors than absolute ages (see \cite{JMDPP04} for detailed
discussion, specially their table~2). The value of $H(z)$ is then
directly computed by using Eq.~\ref{eq:hz}. This is shown in
fig.~\ref{fig:agevsz} with $1\sigma$ error bars. Also shown (dotted
line) is $H(z)$ for the LCDM model. 

\subsection{Constraints on the potential}

Following the discussion in sec \ref{sec.V_cheby}, we present constraints on
the shape of the potential achieavable from present and future data sets. 
Figure (\ref{fig:MCMC_pot_ages}) shows the  constraints on the first
three Chebyshev coefficient for the potential that can be obtained
from our galaxy sample, combined with the determination of the Hubble constant at $z
=0.09$ obtained by \cite{JVTS03} from the SDSS luminous red galaxies.
We have assumed a flat Universe and marginalized over 
 a gaussian prior  on $\Omega_{m,0}$  ($\Omega_{m,0}=0.27\pm 0.07$
 (e.g.,\cite{Verde+2dF02}) and a flat prior on $H_0$
 ($30<H_0<100$ Km/s/Mpc). We have used only the large scale structure
 prior on $\Omega_{m,0}$, as the determination of \cite{Verde+2dF02} is
 insensitive to dark energy. Conversely, CMB constraints on the matter
 density of the Universe are highly sensitive to the assumptions about
 the nature of dark energy (see e.g., \cite{SpergelWMAP03} in
 particular figure 12), and thus should not be used
 in this context. Of course, the addition of CMB  data can greatly
 improve the constraints on the nature of dark energy, but this need to
 be done in a joint analysis and it is left to future work.
 
\begin{figure}
\includegraphics[scale=1.0]{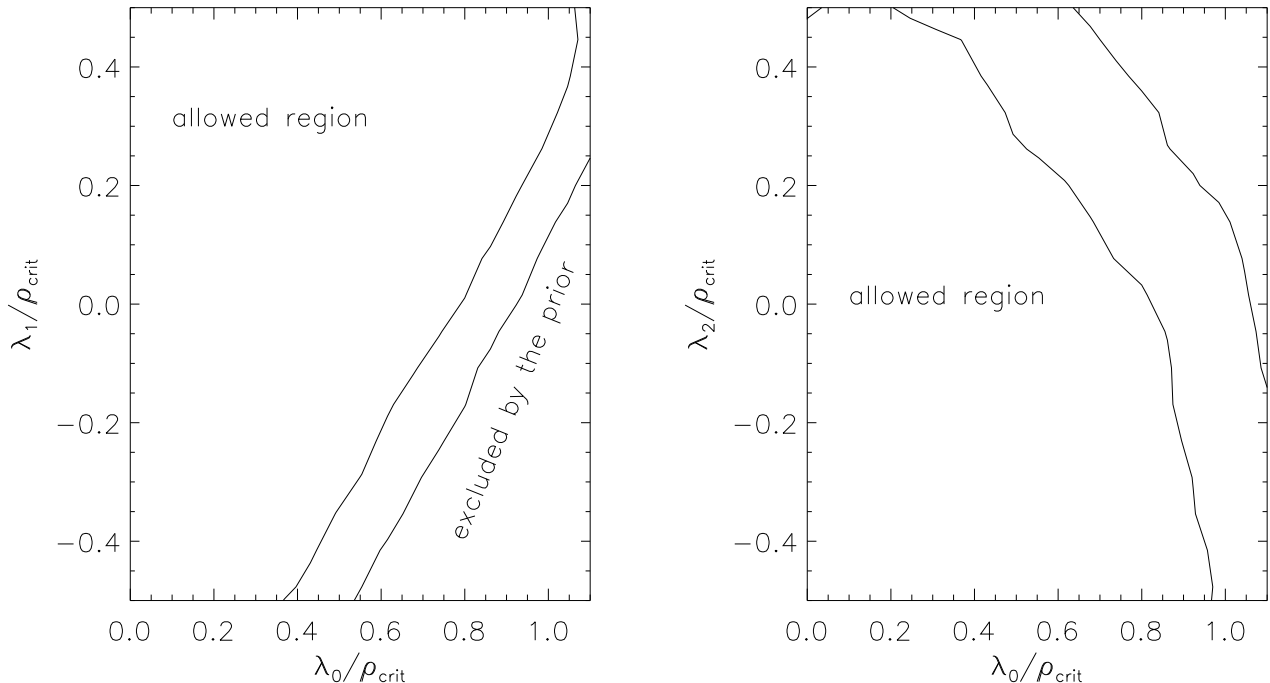}
\caption{Regions in the  $\lambda_1/\rho_{c}$ vs
  $\lambda_0/\rho_{c}$ (left panel) and $\lambda_2/\rho_{c}$ vs
  $\lambda_0/\rho_{c}$ (right panel)  excluded  at the $1-\sigma$ and $2-\sigma$ joint confidence level, by the priors and the constraints that the kinetic energy in the
  quintessence field  must be positive and that at all redshifts $\rho_m+\rho_q$ must be positive.}
\label{fig.MCMCnodata}
\end{figure}

\begin{figure}
\includegraphics[scale=1.0]{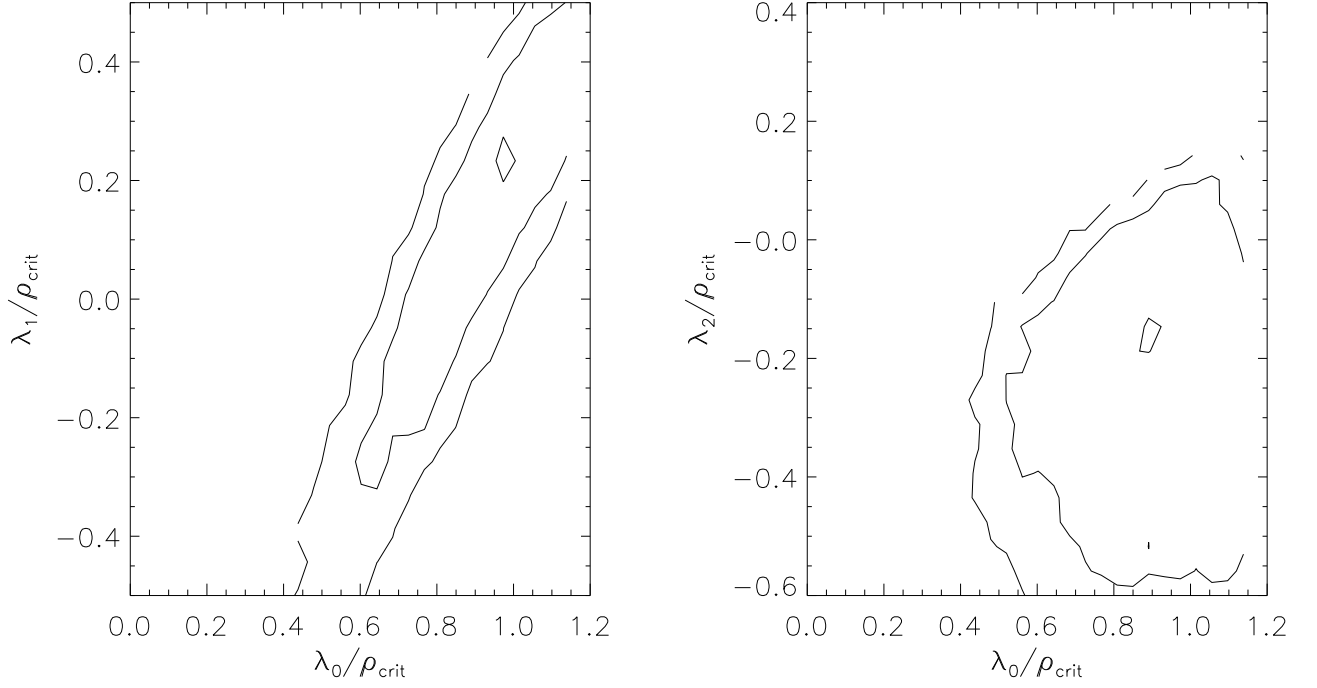}
\caption{Constraints in the $\lambda_1/\rho_{c}$ vs
  $\lambda_0/\rho_{c}$ (left panel) and $\lambda_2/\rho_{c}$ vs
  $\lambda_0/\rho_{c}$ (righ panel) obtained from $H(z)$ measurement based on relative galaxy ages.  Contour levels are 1$-\sigma$
  marginalized, 1$-\sigma$ joint and  2$-\sigma$ joint. The diamon
  shows the location of the maximum of the marginalized likelihood.} 
\label{fig:MCMC_pot_ages}
\end{figure}

Some regions of the parameter space are unphysical as they would
yield a negative kinetic energy or $\rho_m+\rho_q<0$; the combined effect of these priors  in the $\lambda_1/\rho_{c}$ vs
  $\lambda_0/\rho_{c}$ plane and  $\lambda_2/\rho_{c}$ vs
  $\lambda_0/\rho_{c}$ plane is shown in fig. \ref{fig.MCMCnodata}. We consider only the region $0<\lambda_0/\rho_c<1.1$, $-0.5<\lambda_1/\rho_c<0.5$ and $-0.5<\lambda_2/\rho_c<0.5$.

\begin{figure}
\includegraphics[scale=1.0]{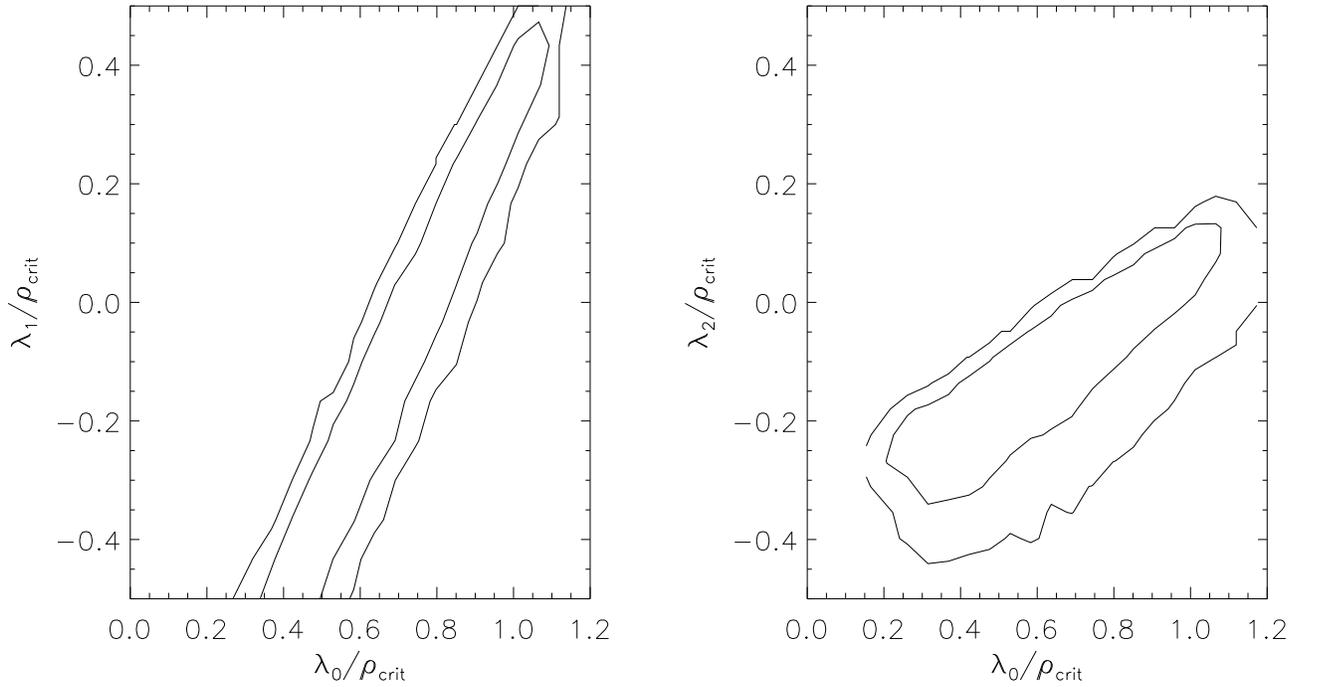}
\caption{One and two
  sigma joint constraints
  in the $\lambda_1/\rho_{c}$ vs
  $\lambda_0/\rho_{c}$ plane and  $\lambda_2/\rho_{c}$ vs
  $\lambda_0/\rho_{c}$ obtained from the Riess et al.  (2004)
  supernovae data. }
\label{fig:MCMC_SN}
\end{figure}

In fig. \ref{fig:MCMC_pot_ages} we show the one and 2 sigma joint
confidence contours in the $\lambda_0/\rho_c$ vs $\lambda_1/\rho_c$ and $\lambda_0/\rho_c$ vs $\lambda_2/\rho_c$ planes, obtained from our $H(z)$ determination. When
adding the HST key project prior on $H_0$  \cite{Freedman+01} the contours remain virtually unchanged.
For comparison in figure \ref{fig:MCMC_SN} we show the constraint
obtained by using the recent supernovae data of \cite{Riess+04}.

\begin{figure}
\includegraphics[scale=0.85]{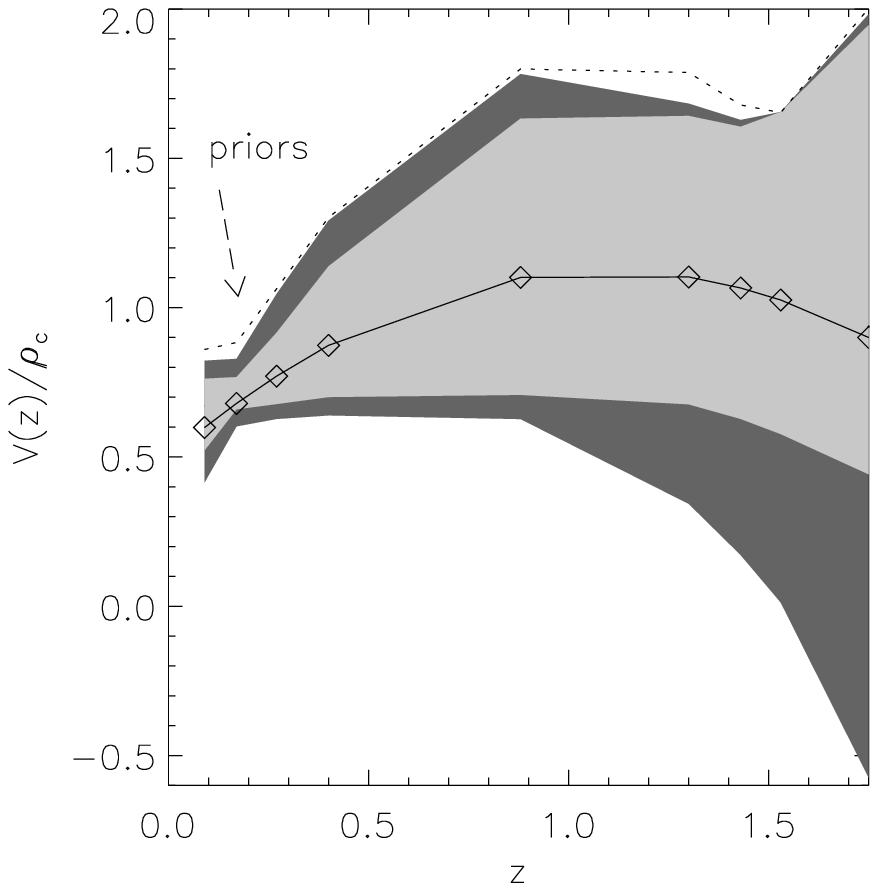}
\includegraphics[scale=0.85]{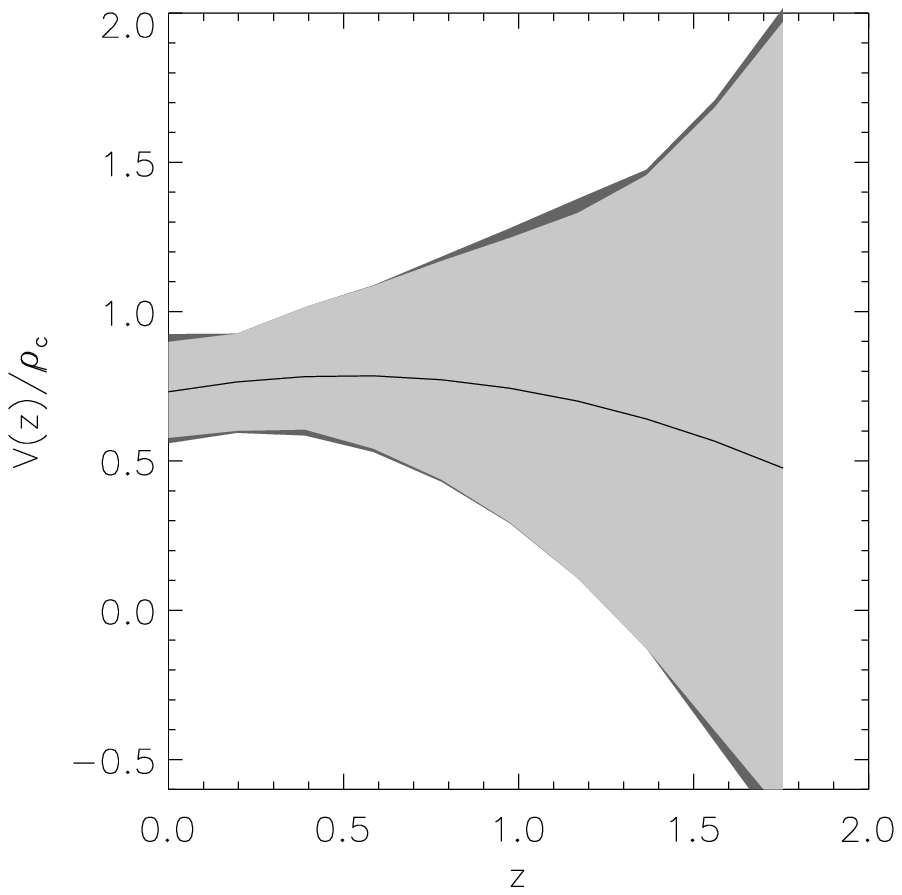}
\caption{Reconstruted $V(z)$ from relative galaxy ages (left) and from
  Supernovae (right). The gray regions represent the 1- and 2-
  $\sigma$ confidence regions. In the left panel the dotted line shows
  the constraint imposed by the prior.}
\label{fig.Vzages}
\end{figure}
Figure \ref{fig.Vzages} shows our best fit  reconstructed $V(z)$ from our $H(z)$ determination (left panel) and from the SN data (right panel), and
the 68\% and 95\% confidence regions.
The present constraints are consistent at the 1-$\sigma$ level with a
constant potential (that is the cosmological constant scenario). 

The two determinations (one based on relative galaxy ages and one SN
data) are consistent with each other. The two methods are completely
independent and are based on different underlying physics, different
assumptions and affected by systematics of completely different nature.
The fact that they agree indicates that possible systematics are
smaller than the statistical errors.

With current data there is a degeneracy between the first two
coefficients, but we can place an upper limit to the kinetic energy in
the quintessence field today: the contribution of the kinetic term to
$\rho_q$ is less than 40\% at the 2-$\sigma$ level and  the best fit
value is at 0.

\begin{figure}
\includegraphics[scale=0.62]{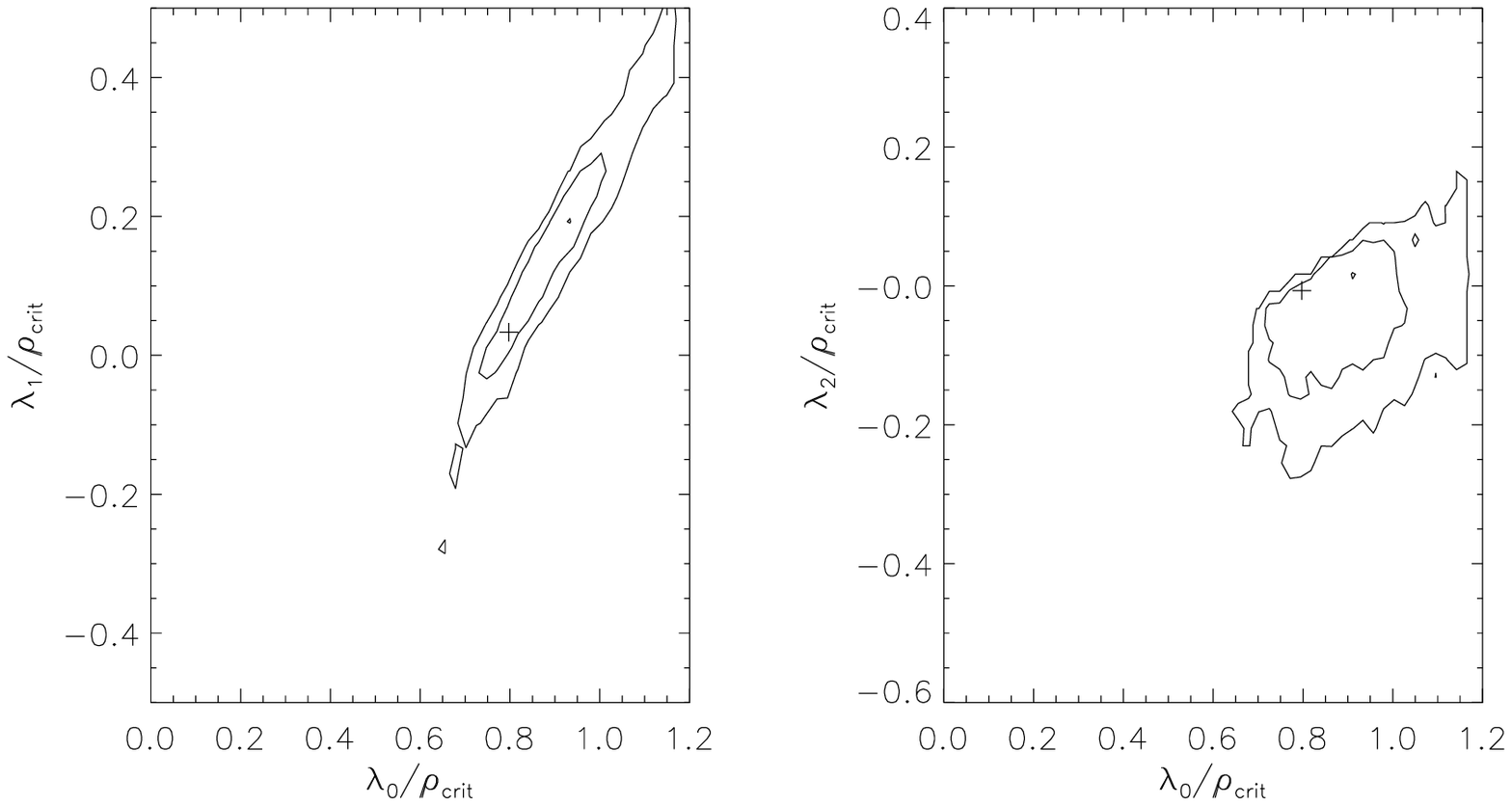}
\includegraphics[scale=0.62]{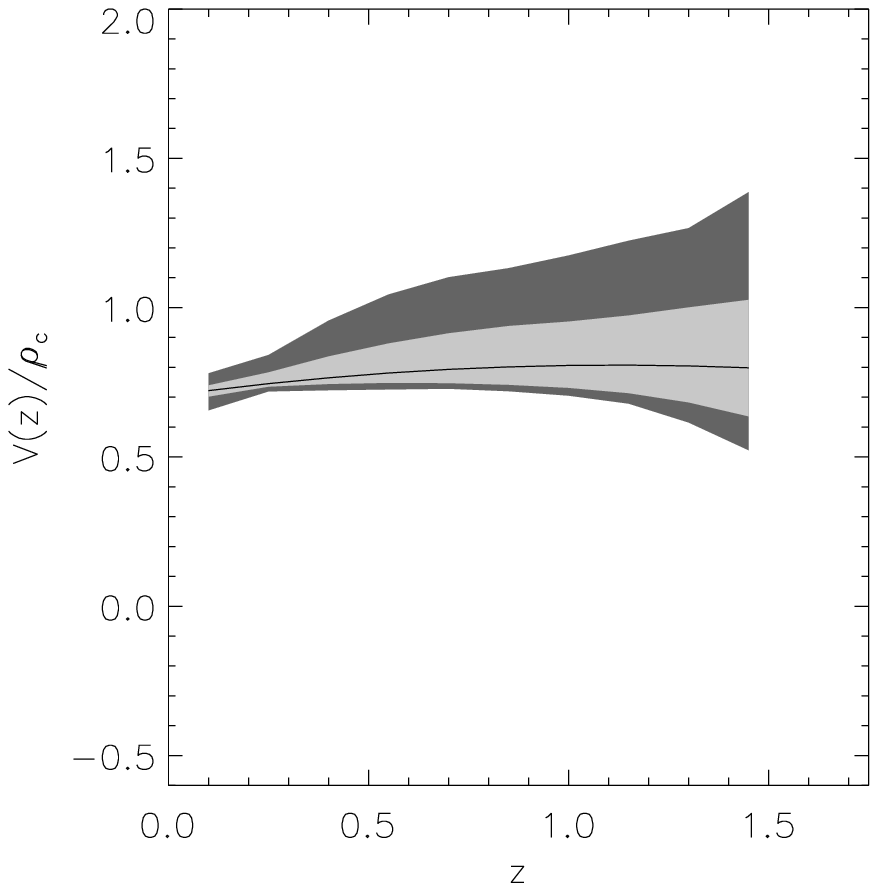}
\caption{Predicted constraints for a experiment with $2000$ galaxies for 
  which ages are measured with an accuracy of $\sim 10$\%. The constraint
  in the Chebyshev coefficients (left panels; circles show the
  location of the maximum marginalized lielihood while '+' show the
  location of the maximum of the joint -5D- likelihood)
   and the reconstructed dark energy potential (right panel) are 
  significantly better than
  current constraints (see text). We have a LCDM model as fiducial.
  The Atacama Cosmology Telescope will identify about 500 galaxy
  clusters in the redshift range $0.1 < z < 1.5$, for at least $2000$
  galaxies there will be spectroscopic follow up and therefore galaxy ages
  can be derived.}
\label{fig.VzagesACT}
\end{figure}

The Atacama Cosmology Telescope (ACT;\cite{KosowskyACT} {\tt
  www.hep.upenn.edu/act}) will identify, through their
Sunyaev-Zeldovich signature in the cosmic microwave background, all
galaxy clusters with masses $> 10^{14}$ M$_{\odot}$ in a patch of the
sky of angular size 100 square degrees. Thus ACT will yield $\gap 500$
galaxy clusters in the redshift range $0.1 < z < 1.5$ . For all these
clusters, spectra of the brightest galaxies in the cluster will be
obtained by South African and Chilean telescopes. This will provide us
with an unbiased sample of $\gap 2000$ passively evolving galaxies
from $z=1.5$ to the present day. To estimate the performance of ACT
galaxies at reconstructing the dark energy potential, we have
estimated that we will have 2000 galaxies for which ages have been
determined with $\sim 10$\% accuracy and therefore $\sim 1000$
determinations of $h$ with $\sim 15\%$ error.  Our\footnote {LV and RJ
  are members of the ACT science team and plan to apply this tecnique
  to ACT data when available in $\sim 2006-2007$} forecasts in the
reconstruction of the dark energy potential are shown in
fig.~\ref{fig.VzagesACT}. We have marginalized over a flat prior on
the Hubble constant $30<H_0<90$ km s$^{-1}$ Mpc$^{-1}$ and a gaussian
prior on $\Omega_{m,0}$, $\Omega_{m,0}=0.27 \pm 0.035$, as an estimate
of the improvement of this determination from galaxy surveys.

\subsection{Constraints on the equation of state}

It is illustrative to work out the consequences of the constraints found
on $\lambda_0, \lambda_1, \lambda_2$. Let's consider the potential
\eqref{eq:ratra-peebles_Vz} that gives rise to a constant   equation of
state. If $\alpha$ is small then one can approximate the potential
with $\lambda (1+\alpha z)$ and thus identify the coefficients in the
Chebyshev expansion: $\lambda_0 \longrightarrow \lambda$ and
$\lambda_1\longrightarrow \lambda\alpha$, $\lambda_2 =0$. 
We thus obtain $\lambda_1/\lambda_0<0.3$ at the 1-$\sigma$
level; since $\lambda_1/\lambda_0 \simeq \alpha =3(w+1)$ we obtain
$w\lap-0.9$ at the 1$-\sigma$ level.

\begin{figure}
\includegraphics[scale=0.8]{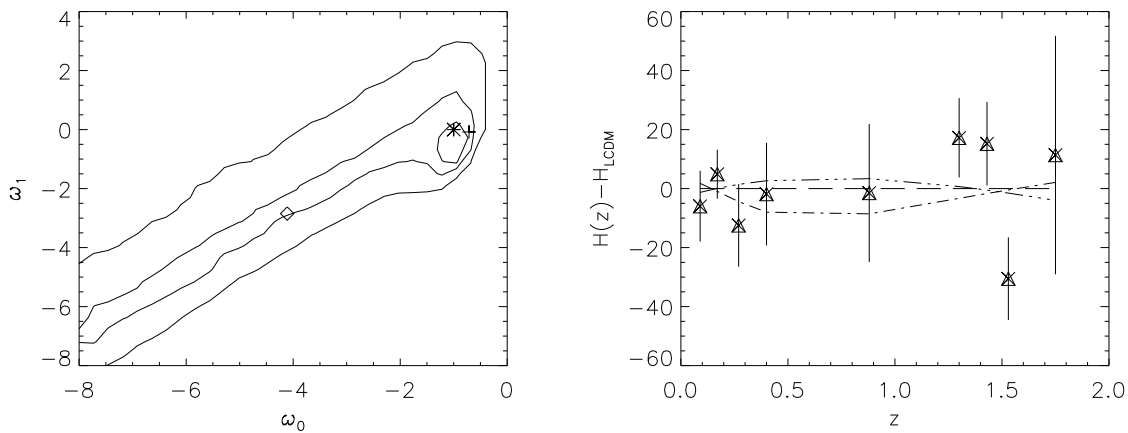}
\caption{Left panel: Constraints in the $\omega_0$ vs $\omega_1$
  obtained from our galaxy sample with the differential ages
  method. The contours show the $1-\sigma$ marginalized, $1-\sigma$
  and $2-\sigma$ joint confidence levels. The degeneracy constrains $w_0\equiv\omega_0-\omega_1$. Three points in the
  parameter space are selected. Right panel: difference between the
  hubble constant in a given model and the Hubble constant in the LCDM
  model. The points with error-bars are our data points, the
  long-dashed line corresponds to the LCDM model (*-point in the left
  panel), the dot-dashed line corresponds to the ``diamond''-point and
  the dot-dot- dot-dashed one to the '+'-point.}
\label{fig:wages}
\end{figure}
As illustrated in sec. \ref{sec.w_cheby} for more general cases 
we can expand the redshift evolution of the equation of state
parameter in terms of Chebyshev polynomials. Here we show how 
constraints on $w(z)$ obtained from our galaxy-sample with the
differential ages method  compare with other constraints.
For example in figure \ref{fig:wages} (left panel) we show the constraints in the
plane $\omega_0$ vs $\omega_1$ (i.e. we impose $N=1$ in
\ref{eq:w_cheby}), 
where we have used the HST key prior for $H_0$ and the prior
$\Omega_{m,0}=0.27\pm 0.04$ as in  \cite{Riess+04}. The contours
show the $1\sigma$ marginalized, $1\sigma$ and $2\sigma$ joint confidence levels.
To compare with the SN constraints of \cite{Riess+04} recall that
their $w_0$ is $\omega_0-\omega_1$. Thus the degeneracy seen in the
figure is a constraint on  $w_0$. Three points in the
$\omega_0$ vs $\omega_1$ parameter space are indicated by the
diamond, star and '+' sign. These points are at the $1\sigma$ joint
confidence level, well within the  $1\sigma$ marginalized level and
at the $1\sigma$ marginalized level, respectively. In particular the
'*' point correspond to the LCDM model.
In the right panel we show the difference between the Hubble parameter
for a given model and that in the LCDM case. Also our determinations
of $H(z)$ are shown. The long-dashed line corresponds to the LCDM case
(* point), the dot-dashed line corresponds to the ``diamond''-point and the
dot-dot-dot-dashed line to the '+'-point. It is clear that more
data-points in the redshift range around $z=0.7$ would help in breaking the degeneracy. 
 
\section{Slow-roll dark energy}

The constraints derived from our observational determination of $H(z)$
combined with our theoretical analysis suggest that observations in
the redshift range $0.1 < z < 1.8$ are consistent, at the $1\sigma$
level, with a cosmological constant equation of state 
$(w =-1)$.

This suggests to analyse more closely the conditions under which a
quintessence field could resemble such an equation of state in that
redshift range, because this is the challenge we will be facing in the near 
future.

There are at least two different approaches that one can attempt:
either work with a generic potential and determine the properties it
has to satisfy to resemble a cosmological constant, or attempt to
argue some universality in the functional form of the potential due to
its expected flatness in field space.

\subsection{Slow-roll in redshift}
\label{eq:geninf}

Given a generic potential scalar field in the presence of a
non-negligible matter energy density $\rho_m(z)$, we would expect the
conditions the potential has to satisfy to be a natural generalisation
of the slow-roll conditions during inflation, including the effects of
matter. These two conditions are :
\begin{equation}
  w_q (z) \approx -1\,,\quad \frac{d w_q(z)}{dt}\approx 0\,\quad
  \forall\,z\in\,[0,\,z_0]~.
 \label{eq:qvsc}
\end{equation}
The first one ensures that dark energy behaves
approximately as a
cosmolgical constant at a given redshift $z$, whereas the second
ensures that such property is maintained in time.

There are several equivalent ways of studying the consequences of
these conditions. In terms of the kinetic scalar field energy $K[q]$ 
and its potential energy $V[q]$, Eq. \eqref{eq:qvsc} implies that 
$V[q]\gg K[q]$ and that the ratio, $K[q]/V[q]$ is nearly constant in time:
\begin{equation}
  \begin{aligned}[m]
    w_q(z)\approx -1 \quad & \Rightarrow \quad \frac{K[q]}{V[q]}\ll
    1~, \\
    \frac{d w_q(z)}{dt}\approx 0 \quad & \Rightarrow \quad
    \frac{d\,K[q]/V[q]}{dt}\approx 0~.
  \end{aligned}
 \label{eq:obscond}
\end{equation}

In terms of the fundamental degrees of freedom, $q(t)$, conditions
\eqref{eq:qvsc} are equivalent to
\begin{eqnarray}
  w_q(z)\approx -1 \quad & \Rightarrow & \quad \frac{1}{2}\dot{q}^2 \ll
  V[q]~, \label{eq:fc1} \\
  \frac{d w_q(z)}{dt}\approx 0 \quad & \Rightarrow & \quad
  \frac{\ddot{q}}{V^\prime[q]}\approx \frac{K[q]}{V[q]} \ll 1~,
 \label{eq:fc2}
\end{eqnarray} 
where the last inequality is derived from the identity
\begin{equation}
  \frac{d w_q}{dt} = 2\,\frac{\dot{q}\,V[q]}{\rho_q^2}\,\left\{
  \ddot{q} - \frac{K[q]}{V[q]}\,V^\prime[q]\right\}~.
\end{equation}  
Under these circumstances, the first Friedmann equation
\eqref{eq:friedmann} and the Klein-Gordon equation \eqref{eq:qmotion}
reduce to
\begin{equation}
  \begin{aligned}[m]
    3\,\frac{H^2}{\kappa} & \approx  \rho_m + V~, \\
    3H\,\dot{q} &\approx  -V^\prime~,
  \end{aligned}
 \label{eq:motionapp}
\end{equation}
which are the extension of the slow-roll equations used in inflation 
in the presence of matter. One can now rewrite conditions
\eqref{eq:fc1}  and  \eqref{eq:fc2} respectively as
\begin{equation}
  \left(\frac{m_p\,V^\prime}{V}\right)^2 \ll 48\pi\,
  \left(1+\frac{\Omega_m}{\Omega_q}\right)~,
 \label{eq:fc1bis}
\end{equation}
\begin{equation}
  m_p^2\,\frac{V^{\prime\prime}}{V} \ll 24\pi\,
  \left(1+\frac{3}{2}\,\frac{\Omega_m}{\Omega_q}\right)~,
 \label{eq:fc2bis}
\end{equation}
where we already used the fact that $\rho_m  V^{-1}\sim
\Omega_m \Omega_q^{-1}$ whenever \eqref{eq:fc1} is satisfied.

Following the discussion in section \ref{sec:redpar}, it is also
convenient to rewrite these conditions in terms of redshift
derivatives of the potential $V[q(z)]$. The
analogue of conditions \eqref{eq:fc1bis} and \eqref{eq:fc2bis} are:
\begin{eqnarray}
  \frac{1}{V}\,\frac{dV}{dz} & \ll & \frac{6}{1+z}~,
 \label{eq:cvz1} \\
  \left(\frac{dV}{dz}\right)^{-1}\,\frac{d^2V}{dz^2} & \ll &
  \frac{5}{1+z}~.
 \label{eq:cvz2}
\end{eqnarray}

\subsection{Slow-roll in the field}
 \label{sec:flat} 

Phenomenologically, there are many inequivalent functionals that could
be chosen to describe the quintessence field dynamics. Each of them,
would typically depend on a set of undetermined parameters, which
would be determined by fitting them to observations, as we did in
Section \ref{sec:exp}. 
In order for a generic potential to look indistingishable from a
cosmological constant, these parameters need to be highly fine
tuned. 

It is precisely this fine tuning that suggests that, {\it
  independently} of the functional form of the potential, the
potential will allow an expansion in terms of the variation of the
unobservable scalar field variation $\Delta_q(t)=q(t)-q(0)$, measuring
its variation from its current value today.

Let's assume that there is a certain period of physical time around today, i.e. $t=0$,
and consistent with the range of redshift covered in this work,
where the variations in the scalar potential are small in field space. In other words,
the potential is ``flat''. Under these conditions, and independently of its functional 
form, the potential $V[q]$ can be approximated by
\begin{equation}
  V[q] \approx V[q(0)] + V^\prime[q(0)]\Delta_q(t) + 
  \frac{1}{2}\,V^{\prime\prime}[q(0)]\left(\Delta_q(t)\right)^2
  + \eO \left((\Delta_q(t)))^3\right)~.
 \label{eq:scheme}
\end{equation}
Let us emphasize that such an expansion is always viable for small
enough $\Delta_q$, but the  Taylor expansion can have a
wider validity if the potential is flat enough, that is if the 
derivatives of the potential are small $|V^n|<<V_0$ and if the kinetic
energy is small $K<<V_0$.

For this to be a good approximation the following two conditions should
be satisfied:
\begin{eqnarray}
  \frac{1}{2}\,V^{\prime\prime}_0\,(q(t)-q(0)) & \ll & V^\prime_0~, 
 \label{eq:app1} \\
  V^\prime_0\,(q(t)-q(0)) & < & V_0~,
 \label{eq:app2}
\end{eqnarray}
where we introduced the notation $V^{(n)}_0 =
\frac{d^n\,V[q]}{d\,q^n}[q(0)]$.

We shall also assume that the
energy of the scalar field $q(t)$ is dominated by the potential energy
\begin{equation}
  \frac{1}{2}\dot{q}^2 \ll V[q]~.
 \label{eq:pdom}
\end{equation}
so that the scalar field dynamics can resemble a cosmological constant (see \eqref{eq:obscond})
and the rolling due to the kinetic energy is small.
In the following, we shall proceed to attempt to integrate the system perturbatively.
At zeroth-order {\bf check} in the potential expansion the first of Friedmann's
equation in \eqref{eq:friedmann} reduces to
\begin{equation*}
  H^2 \sim \frac{\kappa}{3}\left(\rho_{T,0} \left(\frac{a_0}{a}\right)^{3(1+w)} +
  V_0\right)~.
\end{equation*}
where $a(0)\equiv 1$.
If $\rho_{T,0}=\rho_{m,0}$, $w=0$ and this reduces to the case
of a LCDM universe.

The exact solution
\begin{equation}
  3H = \frac{2c}{1+w}\,\frac{y_0+\tanh\,ct}{1+y_0\,\tanh\,ct}~,
 \label{eq:flathubble}
\end{equation}
and 
\begin{equation}
  \left(a(t)\right)^{3(1+w)} = (\cosh\,ct)^2\,(1+y_0\,\tanh\,ct)^2~,
 \label{eq:sfactor}
\end{equation}

becomes, for matter + dark energy universe 
\begin{equation}
  a(t)^3 = (y_0^2-1)\,\sinh^2\,(ct+\hat{c})~,
 \label{eq:sfactor1}
\end{equation}
which yields a Hubble parameter, at zeroth order $H(t)^{-1}=3\,\tanh
(ct+\hat{c})/(2c)$, where we have introduced two dimensionless
parameters:

\begin{equation}
  y_0 = \sqrt{\frac{\rho_0}{V_0} + 1}~, \quad
  \tanh\,\hat{c} = y_0^{-1}~,
\end{equation}
and the dimensional one
\begin{equation}
  c = \sqrt{3\kappa\,V_0}\frac{1+w}{2}~.
\end{equation}
In a matter + dark energy universe $(w=0)$ and in our current approximation 
$c\sim 3H_0\sqrt{\Omega_{q,0}}/2\sim 0.09\, Gyr^{-1}$,
whereas $y_0 \sim \sqrt{\Omega_{m,0}/\Omega_{q,0} +1}\sim 1.18$ and
$\hat{c}\sim 1.2$. Thus the  age of the Universe is $\sim$ 13 Gyrs.

We can then proceed to integrate
the Klein-Gordon equation by taking the first non-trivial contribution coming from the expansion
\eqref{eq:scheme}, and plugging in the zeroth order Hubble parameter
\begin{equation}
  \ddot{q} + 3H\,\dot{q} + V^\prime_0 \sim 0~.
\end{equation}
The solution for the velocity of the scalar filed is
\begin{equation}
\dot{q}=\frac{k-F(t)}{(\cosh ct +y_0 \sinh ct)^{2/(1+w)}}
\end{equation}
where
\begin{equation}
F(t)=V_0'\int_0^t\left[\cosh cx+y_0 \sinh
  cx\right]^{2/(1+w)}dx=(\mbox{ if $w=0$})\,\,
  \frac{1}{2c}\,V^\prime_0\,(y_0^2-1)\,
  \left(\frac{1}{2}\,\sinh\,2(ct+\hat{c}) - (ct+\hat{c})\right)~.
\end{equation}
Here, $k$ is the integration constant and we have used the fact that
$y_0^2-1>0$ and the identity
\begin{equation}
  \cosh\,ct + y_0\,\sinh\,ct = \sqrt{y_0^2-1}\,\sinh\,(ct+\hat{c})~.
\end{equation}

\begin{figure}
\includegraphics[scale=0.7]{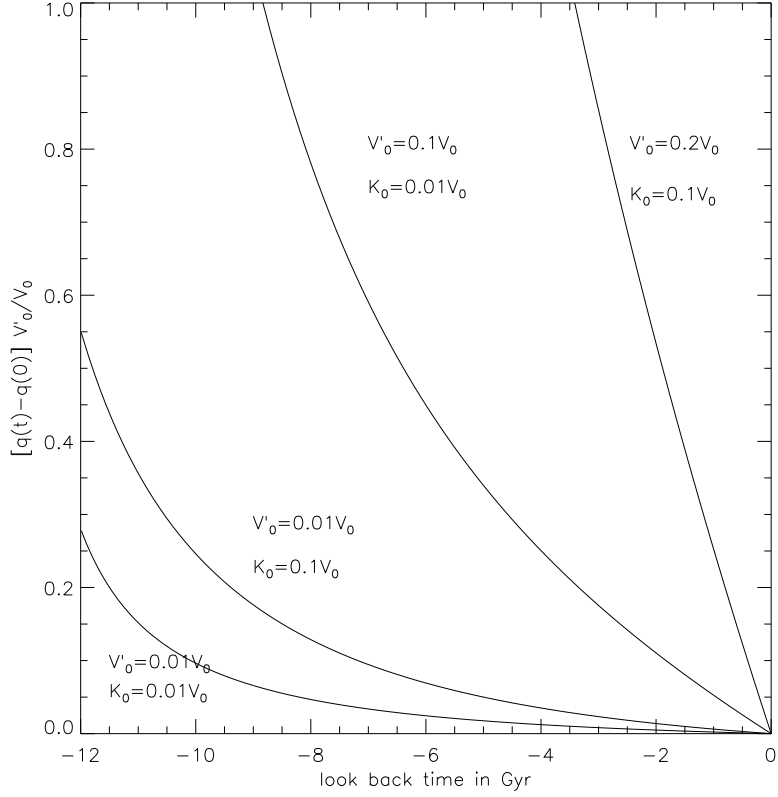}
\caption{Interval in time where the slow roll approximation is valid
  for some choices of  $V_0'$ and $K_0$, when $V_0/\rho_c=0.7$. }
\label{fig:slowroll}
\end{figure}

Note that we could identify the constant of integration $k$ with  the expression
$k=\dot{q}(0) + F(0)$ involving the kinetic energy of the scalar field today.
A second further integration yields the dynamical evolution for the
field,
\begin{equation}
  q(t)-q(0) = k_1 + k\,\int^t\,(a(x))^{-3}\,dx -
  \int^t\,F(x)\,(a(x))^{-3}\,dx~.
 \label{eq:flatq}
\end{equation}
which in a matter + dark energy universe becomes
\begin{equation}
  q(t)-q(0) = \frac{k}{c\,(y_0^2-1)\,\tanh\,\hat{c}\,}\,
  \left(1-\frac{\tanh\,\hat{c}}{\tanh\,(ct+\hat{c})}\right)
  + \frac{V^\prime_0}{2c^2}\,\left(\frac{\hat{c}}{\tanh\,\hat{c}} - 
  \frac{ct+\hat{c}}{\tanh\,(ct+\hat{c})}\right)~.
\end{equation}
where 
\begin{equation}
  k = \dot{q}(0) + F(0) \quad , \quad
  F(0) = (y_0^2-1)\,\frac{V^\prime_0\,\hat{c}}{2c}\,
  \left(\frac{\sinh\,2\hat{c}}{2\hat{c}} - 1\right)~.
\end{equation}

The approximations presented here are therefore valid if
$V_0'/V_0[q(t)-q(0)]<1$. Figure \ref{fig:slowroll} shows, for some
choices of $V0$, $V_0'$ and $K_0$,  the range in lookback time where
this approximation is valid.

\section{Conclusions}
\label{sec:conc}

We have proposed to constrain the nature of dark energy, a rolling
scalar field or a cosmological constant, by
reconstructing its  potential as a function of redshift.
We have presented a formalism, similar to the horizon-flow parameters
in inflation, to relate quantities characterizing the
dark energy dynamics,  i.e. potential 
and kinetic energy densities, to direct observables such as the matter
density $\rho_m (z)$, the Hubble parameter $H(z)$ and its derivatives. 
This is the core of our reconstruction programme, and our 
results are summarised by \eqref{eq:potential}, \eqref{eq:kinetic} and \eqref{eq:vprime}, 
which provide the value of the potential and kinetic energy densities, and the first derivative
of the potential, as a function of redshift. These expressions are valid even in the presence
of higher order curvature corrections to General Relativity and exotic
matter sources. 
In principle, integrating the exact reconstruction formula for the kinetic
energy, allows one to determine the function $q(z)$. Using the latter,
one can infer the real shape of the potential $V[q]$ from the
determination of $V(z)$.

We have then focused on the case
of a expanding Universe with only matter and dark energy
components and at $z\ll 1000$. In this case the above expressions
simplify to \eqref{eq:rpotential}, \eqref{eq.rkinetic} and
\eqref{eq:rvprime}. This exact reconstruction formulas are currently difficult to be used due to the experimental challenges in 
determining the derivatives of the Hubble parameter \ref{sec:exp}.
However, given a 
parameterisation for the potential energy density, the relation \eqref{eq:Hz} becomes a differential equation
for the Hubble parameter which can be integrated analytically. Thus
determinations of $H(z)$ can be used to constrain $V(z)$ directly.
Since effectively one will always be dealing with observations
covering a finite redshift range, by an appropiate linear transformation in the redshift variable, 
we can always work in the interval $[-1,\,1]$, where we know the set of Chebyshev polynomials provide 
a complete orthonormal set of functions, i.e. any function in the interval can be expressed as a linear
combination of Chebyshev polynomials. Moreover, these approximating polynomials have the smallest maximum 
deviation from the true function at any given order, and provide a well-defined estimate of the error introduced
in the truncation of the expansion at a finite order. We point out, in passing, that such a parameterisation
could be used in any other attempts considered in the literature where it was the equation of state the observable
being parameterised by its redshift dependence.

Using  observations of passively evolving galaxies we obtain
measurements of the  Hubble parameter at 9 different redshifts. We use
these determinations to constrain the first three  coefficients in the
Chebyshev expansion of the potential. For comparison we repeat the
analysis using recent supernovae data, which give us an integral of
the Hubble parameter. We find that the reconstructed potentials from
both data sets are consistent,  giving some confidence that  the
results are not heavily plagued by systematic errors.
The standard LCDM model is consistent with current data at the 1$sigma$
level.
We show that future data obtained from the Atacama Cosmology Telescope
will greatly improve the constraints.

Since a cosmological constant  is a good fit to the
observations we asked the question of how to generically describe
small deviations from this simple scenario. 
It is clear that even if the nature of dark energy might be
related to a dynamical field,  distinguishing such a scenario from a
real cosmological constant will be  an  extraordinary
experimental challenge, as the dark energy potential can be
arbitrarily close to a constant.

We thus analysed the conditions for the dark energy field to
``slow-roll'' in the presence of matter, thus enabling a dynamical
dark energy to get arbitrarily close to a cosmological constant \eqref{eq:qvsc}. 
By expanding the potential in Taylor series for  
$\Delta_q(t) = q(t)- q(0)$ we derived 
the generalisation of the standard slow-roll conditions used in inflation, in the presence of matter, which
translate into conditions that the functional $V[q]$ must satisfy to
explain the maximum deviation  allowed from a cosmological constant.

Even though in this paper we focused on single canonically normalised
scalar field, it should be clear that it is straightforward to  apply our formalism to an arbitrary number of them, not necessarily being canonically
normalised. However, the observables quantities  are the matter density, the Hubble
parameter and its derivatives, which depend on the full kinetic and
potential energy densities of the scalar field sector, and are
insensitive to whether these values are given by the superposition of
more than one field.
This opens up the question of whether one would be able to determine, experimentally,
the existence of more than one rolling scalar field. In other words, to which extent it is possible to disentagle the full
kinetic/potential energy of a superposition of scalar fields into the kinetic/potential energies of their components?

We have illustrated the enormous experimental  challenges of reconstructing $V[q]$ from $V(z)$ just for a single scalar field. 
The task is even harder for more than one field as there are more
derivative directions to consider,  $\partial V/\partial q_i$, and
direct experimental observables depend only on the time derivatives of
the full potential energy, i.e.$dV/dt = \partial V/\partial
q_i\,\,\dot{q}_i$.
Such a disentanglement  seems extremely challenging, if not
impossible, at least from the perspective of the formalism developed
here. In this context we can say that our formalism  enables one to reconstruct the
properties (potential and kinetic energy),  of an ``effective'' field.

\section*{Appendix A}

Chebyshev polynomials can be computed using the recursion relation:
\begin{equation}
T_{n+1}(x)=2xT_n(x)-T_{n-1}(x)\;\;\;{\rm for} \;\;\; n\ge 1
\end{equation}
where 

\begin{equation}
T_0(x)=1\;, \;\;\;\; T_1(x)=x
\end{equation}

Thus $T_n(x)$ will have the following structure: $T_n(x)=
\alpha_0 x^{n}+\alpha_2 x^{n-2}+\alpha_4 x^{n-4}+....$,, so for
example:
\begin{eqnarray}
T_2(x)&=&2x^2-1  \\ \nonumber
T_3(x)&=&4x^3-3x  \\ \nonumber
T_4(x)&=&8x^4-8x^2+1 \\
\end{eqnarray}

The integral on the RHS of Eq. \eqref{eq:chebxzpn}   will involve
a series of integrals of the type:
\begin{equation}
I_n=\int_{-1}^{2z/z_{max}-1} x^n (a+bx)^{-7}\,dx
\end{equation}

In particular, using the recursion relation of the Chebyshev
 polynomials we find that: 

\begin{equation*}
  T_0(x)=1 \quad \Rightarrow \quad F_0(x) \equiv
  I_0~
\end{equation*}
\begin{equation*}
  T_1(x)=x \quad \Rightarrow \quad F_1(x)\equiv
 I_1~.
\end{equation*}
\begin{equation*}
  T_2(x)=2x^2-1 \quad \Rightarrow \quad F_2(x)\equiv
  \left[2I_2-I_0\right]~.
\end{equation*}
\begin{equation*}
  T_3(x)=4x^3-3x \quad \Rightarrow \quad F_3(x)\equiv
  \left[4I_3-3I_1\right]~.
\end{equation*}
\begin{equation*}
  T_4(x)=8x^4-8x^2+1 \quad \Rightarrow \quad F_4(x)\equiv
  \left[8I_4-8I_2+I_0\right]~.
\end{equation*}
\begin{equation*}
  T_5(x)=16x^5-20x^3+5x \quad \Rightarrow \quad  F_5(x)\equiv
  \left[16I_5-20I_3+5I_1\right]~.
\end{equation*}
\begin{equation*}
  T_6(x)=32x^6-48x^4+18x^2-1 \quad \Rightarrow \quad F_6(x)\equiv
 \left[32I_6-48I_4+18I_2-I_0\right]~.
\end{equation*}

where
\begin{eqnarray}
I_n&=&
\left(\frac{2z}{z_{max}}-1\right)^{n+1}\left[\frac{1}{6}\frac{(1+z)^{-6}}{a}-\frac{(n-5)(1+z)^{-5}}{30
  a^2}+\frac{(n-5)(n-4)(1+z)^{-4}}{120 a^3}\right.\\
 &-& \left.\frac{(n-5)(n-4)(n-3)(1+z)^{-3}}{360
    a^4}+\frac{(n-5)\cdot \cdot \cdot(n-2)(1+z)^{-2}}{720
    a^5}-\frac{(n-5)\cdot \cdot \cdot (n-1)(1+z)^{-1}}{720
    a^6}\right]\\
&-&(-1)^{n+1}\left[\frac{1}{6a}-\frac{(n-5)}{30 a^2}+\frac{(n-5)(n-4)}{120
    a^3}-\frac{(n-5)(n-4)(n-3)}{360 a^4}
  +\frac{(n-5)\cdot \cdot \cdot(n-2)}{720 a^5}\right.\\
&-&\left.\frac{(n-5)\cdot \cdot \cdot (n-1)}{720
    a^6}\right]+ \frac{(n-5)\cdot \cdot \cdot n}{720 a^6}J_n\\
\end{eqnarray}
and
\begin{equation}
J_n=\int_{-1}^{2z/z_{max}-1}\frac{x^n}{(a+bx)}dx
\end{equation}

These integrals are given by 
\begin{equation}
  J_n= \sum_{m=0}^{n-1}(-1)^m a^m \frac{\left[(2z/z_{max}-1)^{(n-m)}
  -(-1)^{(n-m)} \right]}{(n-m)b^{(m+1)}} + (-1)^n\frac{a^n}{b^{(n+1)}}\log (1+z)
\end{equation}
but can also be obtained using the recursion relation
\begin{equation}
J_n=\frac{1}{nb}\left[(2z/z_{max}-1)^n-(-1)^n\right]-\frac{a}{b}J_{n-1}
\end{equation}
where
\begin{equation}
J_0=\frac{1}{b}\log (1+z)
\end{equation}

Integrals $I_n$ can also be obtained 
by recursive relation \cite{Gradshteyn}
\begin{equation}
  I_n= \frac{(1+z)^{-6}(2z/z_{max}-1)^n-(-1)^n}{(n-6)z_{max}/2}
  -\frac{n\,(1+z_{max}/2)}{(n-6)z_{max}/2}I_{n-1}  \,\,\,\,\,\, \mbox{for $n \ne 6$}
\end{equation}
where 
\begin{equation}
I_0(x)=-\frac{1}{3 z_{max}}\left(\frac{1}{(1+z)^6}-1\right)\,.
\end{equation}

In the case of Eq. \eqref{eq:Hzwcheby} $G_n$ are defined as:
\begin{equation*}
  T_0(x)=1 \quad \Rightarrow \quad G_0(x) \equiv
  J_0~
\end{equation*}
\begin{equation*}
  T_1(x)=x \quad \Rightarrow \quad G_1(x)\equiv
  J_1~.
\end{equation*}
\begin{equation*}
  T_2(x)=2x^2-1 \quad \Rightarrow \quad G_2(x)\equiv
  [2J_2-J_0].
\end{equation*}
etc.

\begin{acknowledgments}
We thank M. Trodden for stimulating discussions. JS would like to thank the 
Institute for Theoretical Physics in Amsterdam, for hospitality during the 
last stages of this project.
The work of RJ is partially supported by NSF grant AST-0206031. 
LV is supported by  NASA grant ADP03-0000-0092.
The work of JS is supported by the DOE under grant DE-FG02-95ER40893 
and by the NSF under grant PHY-0331728.
\end{acknowledgments}


\end{document}